\documentclass[english,aps,prb,twocolumn,footinbib,longbibliography,superscriptaddress,floatfix]{revtex4-1}

\usepackage[english]{babel}
\usepackage{amssymb,amsmath,bbm}
\usepackage[pdftex]{graphicx}
\usepackage{color}
\usepackage[usenames,dvipsnames]{xcolor}
\usepackage{hyperref}
\hypersetup{%
	linktoc=page,%
	colorlinks=true,%
	linkcolor=RoyalBlue,%
	citecolor=OliveGreen,%
	urlcolor=OliveGreen,%
	linkbordercolor=red,%
	pdfborderstyle={/S/U/W 1},%
}
\usepackage{float}
\usepackage{setspace}
\usepackage{amsmath,amssymb}
\usepackage{graphicx}
\usepackage{feynmf}
\usepackage{color}
\usepackage{dsfont}

\newcommand{\rs}{\rm \scriptscriptstyle}

\begin{document}

\title{Time-reversal of an unknown quantum state}

\author{A. V. Lebedev}
\affiliation{Moscow Institute of Physics and Technology, Institutskii per. 9, Dolgoprudny, 141700, Moscow District, Russia}

\author{V. M. Vinokur}
\affiliation{Materials Science Division, Argonne National Laboratory, 9700 S. Cass Av., Argonne, IL 60637, USA}
\affiliation{Consortium for Advanced Science and Engineering (CASE) University of Chicago, 5801 S Ellis Ave, Chicago, IL 60637, USA}

\begin{abstract}
For decades, researchers have sought to understand how the irreversibility 	of the surrounding world emerges from the seemingly time symmetric,
fundamental laws of physics. Quantum mechanics conjectured a clue that final irreversibility is set by the measurement procedure and that the  time reversal requires complex conjugation of the wave function,  which is overly complex to spontaneously appear in nature. Building  on this Landau-Wigner conjecture, it became possible to demonstrate that time reversal is exponentially improbable in a virgin nature and to design an algorithm artificially reversing a time arrow for a given quantum 	state on the IBM quantum computer. However, the implemented arrow-of-time reversal embraced only the known states initially disentangled from the thermodynamic reservoir. Here we develop a procedure for reversing the temporal evolution of an arbitrary unknown quantum state. This opens the route for general universal  algorithms sending temporal evolution of an arbitrary system backwards in time.
\end{abstract}

\maketitle

An origin of the arrow of time, the concept coined for expressing one-way direction of time is inextricably associated with the Second Law of Thermodynamics\,\cite{Lloyd2013}, which declares that entropy growth stems from the system’s energy dissipation to the environment\,\cite{Kelvin,Maxwell,Boltzmann:1872,Boltzmann:1896,Lebowitz:1999}. Thermodynamic considerations\,\cite{Holster:2003,Andrieux2007,Parrondo2009,Campisi2011-1,Campisi2011-2,Deffner2011,Oreshkov:2015,Manzano2017,Santos2017,Batalhao2018-1,Gherardini2018}, combined with the quantum mechanical hypothesis that irreversibility of the evolution of the physical system is related to measurement procedure\,\cite{Landau1927,Neumann1929} and to the necessity of the anti-unitary complex conjugation of the wave function of the system for the time-reversal\,\cite{Wigner1932} led to understanding  that the energy dissipation can be treated in terms of the system's entanglement with the environment\,\cite{Partovi2008,Jennings2010-2,Lloyd2013,Batalhao2015,Camati2016}. The quantum mechanical approach to the origin of the entropy growth problem was crowned by finding that in a quantum system initially not correlated with an environment, the local violation of the Second Law can occur\,\cite{Lesovik2016}. Extending then the solely quantum viewpoint on the arrow of time and elaborating on the implications of the Landau-Neumann-Wigner hypothesis\,\cite{Landau1927,Neumann1929,Wigner1932}, enabled to quantify the complexity of reversing the evolution of the known quantum state and realize the reversal of the arrow of time on the IBM quantum computer\,\cite{us}.

In all these past studies, a thermodynamic reservoir at finite temperatures has been appearing as a high-entropy stochastic bath thermalizing a given quantum system and increasing thus its thermal disorder, hence entropy. Here we demonstrate that most unexpectedly, it is exactly the presence of the reservoir acting in a concert with an auxiliary quantum system with the identical Hamiltonian, $H$, enables to devise the operator of the backward-time evolution $U=\exp(iHt)$ reversing the temporal dynamics of the given quantum system. Importantly, we need not to know the quantum state of this system in order to implement the arrow of time reversal. A dramatic qualitative advance of new protocol  is that it eliminates the need of keeping an exponentially huge record of classical information about the values of the state amplitudes. Moreover, the crucial step compared to the protocol of time reversal of the \textit{known} quantum state\,\cite{us} is that we now lift the requirement that initially the evolving quantum system must be a pure uncorrelated state. Here, we develop a procedure where the initial state can be a mixed state and, therefore, include correlations due to system's past interaction with the environment.

The necessary requirement is that the dynamics of the both, auxiliary and given, systems were governed by the same Hamiltonian $H$. The time reversal protocol comprises the cyclic sequential process of quantum computation on the combined auxiliary and the given systems and the thermalization process of the auxiliary system. A universal time-reversal procedure of an unknown quantum state defined through the density matrix $\hat\rho(t)$ of a quantum system ${\cal S}$ will be described as a reversal the temporal system evolution $\hat\rho(t) = \exp(-i\hat{H} t/\hbar) \hat\rho (0) \exp(+i\hat{H} t/\hbar)$ returning it to system's to original state $\hat\rho(0)$. The only required prior information about the system is its Hamiltonian $\hat{H}$ which is assumed to be completely known. This task is to be accomplished by constructing the backward-evolution unitary operator $\hat{U}^\dagger(t) = \exp(+i\hat{H} t/\hbar)$, and this, in principle, can be done with the help of the universal quantum computer. We first describe how such a procedure can be implemented in a universal manner and estimate its computational complexity. Next, we outline a somewhat more resource-demanding procedure, where, however, one can relax the need of knowing the Hamiltonian $\hat{H}$. We show that if in addition to the quantum system ${\cal S}$ one is provided by an auxiliary system ${\cal A}$, so that $\dim{\cal S} = \dim{\cal A}$, whose dynamics is governed by the same Hamiltonian $\hat{H}$, one can devise $\hat{U}^\dagger(t)$ without knowing an exact form of $\hat{H}$. Finally, we discuss how the partial knowledge on the state $\hat\rho(t)$ can reduce and optimize the complexity of the time-reversal procedure.

The starting point of the reversal procedure is drawn from the observation of S.\,Lloyd and co-authors, see Ref.\,[\onlinecite{Lloyd:2014}] that having an ancilla system in a state $\hat\sigma$ one can approximately construct a unitary operation $\exp(-i\omega\hat\sigma \delta t)$ acting on a system ${\cal S}$ simulating its evolution under Hamiltonian $\hat{H}_a = \hbar\omega \hat\sigma$ during the infinitesimal time interval $\delta t$. Here, $\omega$ refers to some arbitrary rate which for a moment we leave unspecified. Having $N$ identical copies of ancillas, one generates a finite time evolution $\rho(t) \to \rho(t+\tau) = e^{-i\omega\tau\hat\sigma} \hat\rho(t) e^{+i\omega\tau \hat\sigma}$ over the time interval $\tau = N\delta t$ with an accuracy $\propto (\omega\tau)^2/N$ (see Appendix\ref{A1}). The first step of the  time-reversal procedure is then constructing the density matrix $\hat\sigma$. Consider the density operator defined by the given finite-dimensional Hamiltonian $\hat{H}$ having the maximal eigenvalue $\epsilon_\mathrm{max}$:
\begin{equation}
      \hat\sigma = \frac1{Z}\Bigl(\mathds{1} \epsilon_\mathrm{max} -\hat{H}\Bigr),
\end{equation}
where $Z = \epsilon_\mathrm{max} \dim{\cal S} - \mbox{Tr}\{\hat{H}\}$ is the normalization factor. Then the Lloyd (LMR) procedure maps the initial density matrix $\hat\rho(t)$ to
\begin{equation}
\hat\rho(t) \to  \exp\Bigl(\frac{i\omega}{Z} \hat{H}\tau\Bigr) \hat\rho(t) \exp\Bigl( -\frac{i\omega}{Z} \hat{H}\tau\Bigr).
\end{equation}
One sees that application of the LMR procedure with the specific density matrix $\sigma$ realizes approximately the time-reversed evolution of the system
\begin{equation}\label{eq:rhoback}
\hat\rho(t) \to \hat\rho\Bigl( t - \frac{\hbar\omega}{Z}\,\tau \Bigr) + \delta\hat\rho(\tau)
\end{equation}
to a backward delay $\tau_R = (\hbar\omega/Z) \tau$. The accuracy $\delta\hat\rho(\tau)$ of such a time-reversal procedure is given by (see Appendix\ref{A1}),
\begin{eqnarray}\label{eq:rev_accuracy}
|| \delta\hat\rho(\tau)|| \leq \frac{(\omega\tau)^2}{N} \,\Bigl( ||\hat\sigma|| +||\hat\rho(t)|| +2 ||\hat\rho(t)||\, ||\hat\sigma||^2 \Bigr),
\end{eqnarray}
where $||\hat{A}||$ is the operator norm: $||\hat{A}|| = \sup_{|\psi\rangle} \sqrt{\langle \psi| \hat{A}|\psi\rangle/\langle \psi|\psi\rangle}$.

From Eqs.(\ref{eq:rhoback}) and (\ref{eq:rev_accuracy}) one draws two important conclusions. First, the above time-reversal procedure for a backward delay $\tau_R$ requires itself the time $\tau$ to be completed. Therefore, while exercising the reversal, the system still maintain the forward evolution governed by its own Hamiltonian.  Taking this into account, one has to modify Eq.\,(\ref{eq:rhoback}) to
\begin{equation}
       \hat\rho(t) \to \hat\rho\left(t-\tau \frac{\hbar\omega}{Z} + \tau\right),
\end{equation}
which immediately poses the constraint on the operation rate $\omega$ of the LMR procedure: the actual time reversal occurs only for $\hbar\omega > Z$. If this constraint is not satisfied, the time-reversal procedure only slows down the forward time evolution of the system. For a quantum system ${\cal S}$, the threshold rate $Z/\hbar$ is proportional to the Hilbert space dimension $\dim{\cal S}$: $Z = \hbar\tilde\omega \dim{\cal S}$ with $\hbar\tilde\omega = \bigl( \epsilon_\mathrm{max} - \mbox{Tr}\{\hat{H}\}/\dim{\cal S}\bigr)$, which is typically an exponentially large number. In particular, in order to make the time-reversal with the same rate as the forward time evolution one has to demand $\omega > 2Z/\hbar$. This brings straightforwardly the second conclusion: as far as $\omega$ is large, the infinitesimal time step $\delta t$ of the procedure has to be small so that $\omega\delta t \ll 1$, therefore the number $N$ has to be large. Indeed, fixing the backward delay $\tau_R$, the operation rate $\omega = 2Z/\hbar$, and setting the reversal accuracy $\epsilon$: $||\delta\hat\rho(\tau=\tau_R)|| \leq \epsilon$ one finds from Eq.(\ref{eq:rev_accuracy}):
\begin{equation}\label{eq:NvsTR}
N_\epsilon = \frac{||\hat\rho(t)||}{\epsilon}\, \Bigl(\dim{\cal S}\frac{\tau_R}{\tilde\tau}\Bigr)^2,
\end{equation}
where $\tilde\tau = \tilde\omega^{-1}$ is the typical time-scale of the system dynamics and $||\hat\sigma|| \propto (\dim{\cal S})^{-1} \ll ||\hat\rho(t)||$ is  assumed. Equation\,(\ref{eq:NvsTR}) implies that the computational complexity of the time-reversal procedure for an unknown quantum state is proportional to the square of the system's Hilbert space dimension. In contrast, the time reversal of a known pure quantum state $\hat\rho(t) = |\psi(t)\rangle \langle \psi(t)|$ is proportional of the dimension of the Hilbert space which is swept by the system in the course of its forward time evolution $|\psi(0)\rangle \to |\psi(t)\rangle$\cite{us}. As follows from Eq.\,(\ref{eq:NvsTR}), the time-reversal computational cost of an unknown pure state is maximal as long as $||\hat\rho||=1$ in this case. For a mixed high-entropy state $\hat\rho$, the reversal complexity is reduced: given a state $\hat\rho$ with the entropy $S_\rho = \ln(\dim{\cal S})-k\ln(2)$ where only $k\ll \log_2(\dim{\cal S})$ bits of information is known, the upper estimate for the reversal complexity is given by (see Appendix\ref{A2})
\begin{equation}\label{eq:hentN}
      N_\epsilon \leq \frac{k}{\epsilon \log_2(\dim{\cal S})}\, \Bigl(\dim{\cal S}\frac{\tau_R}{\tilde\tau}\Bigr)^2.
\end{equation}

Having a complete information about the Hamiltonian $\hat{H}$ allows one to construct a corresponding quantum circuit realizing the forward time evolution operator $\hat{U} = \exp(-i\hat{H}t/\hbar)$ through a specific fixed set ${\cal G}$ of universal quantum gates: $\hat{U} = \hat{U}_1 \cdots \hat{U}_N$, $\hat{U}_i \in {\cal G}$. As far as ${\cal G}$ is an universal set, for every $\hat{U}_i\in {\cal G}$ one can construct the inverse gate $\hat{U}^\dagger_i$. Therefore, the time-reversed evolution operator $\hat{U}^\dagger$ can be constructed in a purely algorithmic way given the gate decomposition of $\hat{U}$. Thus, the above procedure may appear an extremely ineffective for a practical the time-reversal task. However, the situation turns completely different if we relax the requirement of the exact knowledge of $\hat{H}$ and assume what one, instead, is provided by the equivalent copy of the system ${\cal S}$ governed by the same Hamiltonian $\hat{H}$.

Let us let  one be equipped with the thermodynamic bath at the temperature $T=\beta^{-1}$ in addition to the ancilla. One can then thermalize the ancilla and prepare it in the equilibrium state $\sigma_\beta = Z_\beta^{-1}\exp(-\beta\hat{H})$ with $Z_\beta = \mbox{Tr}\{\exp(-\beta\hat{H})\}$ being a statistical sum. For high-enough temperature $\beta \to 0$, one has $\beta\epsilon_\mathrm{max} \sim 1$ and, therefore, $\sigma_\beta \approx Z_\beta^{-1}(1-\beta\hat{H})$ which gives the desired state of the ancilla to implement the reverse evolution through the LMR procedure. In this case
\begin{equation} \label{eq:thermal-reversal}
\hat\rho(t) \to \hat\rho\Bigl(t - \frac{\hbar\omega\beta}{Z_\beta}\, \tau + \tau \Bigr) + \delta\hat\rho(\tau).
\end{equation}
As can be seen from the above equation the actual time reversal requires the operation rate of the LMR procedure to exceed the threshold
\begin{equation}
\omega > \omega_\mathrm{th}=\frac{T}{\hbar}Z_\beta \approx \frac{T}{\hbar} \dim{\cal S}.
\end{equation}
The approximation error $\delta\hat\rho$ splits now to two contributions, $\delta\hat\rho = \delta\hat\rho_1 + \delta\hat\rho_2$, where $\delta\hat\rho_1$ is the approximation error resulting from the LMR procedure, see Eq.(\ref{eq:rev_accuracy}) with $\hat\sigma\to \hat\sigma_\beta$, while the error $\delta\hat\rho_2$ describes the error due to the $\beta$ expansion of the thermal state
$\hat\sigma_\beta$. Assuming $\omega = 2\omega_\mathrm{th}$, i.e. the backward evolution goes with the same rate as the forward
time evolution one finds
\begin{equation}
\delta\hat\rho_2(\tau) = -i \frac{\tau \beta}{\hbar}\, \Bigl[ \hat{H}^2,\hat\rho\bigl(t- \tau\bigr) \Bigr].
\end{equation}
Then for $||\hat\sigma_\beta|| \ll ||\hat\rho(t)||$ one can estimate the net error as
\begin{equation} \label{eq:thermal-accuracy}
|| \delta\hat\rho(\tau)|| \leq \Bigl(4\frac{Z_\beta^2}{N}\, \Bigl( \frac{\tau}{\tau_\beta} \Bigr)^2 + \frac{\tau}{\tau_\beta} (\beta\epsilon_\mathrm{max})^2 \Bigr) || \hat\rho(t-\tau)||.
\end{equation}
where $\tau_\beta = \hbar\beta$. The temperature dependence of two error contributions in Eq.(\ref{eq:thermal-accuracy}) oppositely depends on the inverse temperature: the error due to thermal expansion (second term) reduces as $\beta\to 0$ while the error due to LMR dynamics (first term) increases with decreasing $\beta$. For a given reverse time delay $\tau$ and the number of LMR iterations $N \gg Z_\beta^2 \approx (\dim{\cal S})^2$, one has an optimal temperature
\begin{equation} \label{eq:temp}
\beta\epsilon_\mathrm{max} = \Biggl( 8\frac{Z_\beta^2}{N}\, \frac{\epsilon_\mathrm{max} \tau}{\hbar} \Biggr)^{1/3},
\end{equation}
and the corresponding net accuracy of the reversal procedure is then given by
\begin{equation} \label{eq:taccuracy}
||\delta\hat\rho(\tau)|| = \epsilon = 3\Biggl( \frac{Z_\beta^2}{N} \Biggr)^{1/3} \Bigl( \frac{\epsilon_\mathrm{max}\tau}{\hbar} \Bigr)^{4/3} ||\hat\rho(t-\tau)||.
\end{equation}
Comparing with the case of the known Hamiltonian time-reversal procedure, see Eq.(\ref{eq:NvsTR}), the reversal complexity here is again proportional to the square of the system's Hilbert space dimension, but, at the same time, has more adverse scaling with the reversal duration and the net accuracy.

The above analysis does not need any prior information about the state $\hat\rho$ which would require very high temperature of the auxilliary thermostat in order to cover all the possible energy states of the system's Hilbert space that finally results in a tremendously high rate $\sim \hbar\beta \dim{\cal S}$ of the LMR procedure.  If, however, some information about the energy content of the state $\hat\rho$ is available, one can appreciably reduce the reversal cost. Indeed, let the state $\hat\rho$ have the average energy $\bar{E} = \mbox{Tr}\{\rho \hat{H}\}$ with an energy variance $(\delta E)^2 = \mbox{Tr}\{\hat\rho(\hat{H} - \bar{E})\}$. Then one can present the density matrix as the result of the low-energy contribution, $\hat\rho_< = \hat{P} \hat\rho \hat{P}/\mbox{Tr}\{\hat{P} \hat\rho\}$ and the high-energy remainder $\hat\rho_> = (1-\hat{P})\hat\rho (1-\hat{P})/\mbox{Tr}\{(1-\hat{P})\hat\rho\}$ where $\hat{P} = \sum_{E< E_\mathrm{max}} |E\rangle \langle E|$ is a projection operator to the subspace with energies below some cut-off energy $E_\mathrm{max} > \bar{E}$: $\hat\rho = (1-\epsilon_{\rs E}) \hat\rho_< + \epsilon_{\rs E} \hat\rho_>$. The additional error due to truncating the system Hilbert space to the low-energy subspace is given by the constant $\epsilon_{\rs E}$ which is a probability for the system to be found in the energy state $E> E_\mathrm{max}$, and, according to the Chebyshev inequality, is bound by
\begin{equation}
      \epsilon_{\rs E} \leq \Bigl( \frac{E_\mathrm{max}-\bar{E}}{\delta E}\Bigr)^{-2}.
\end{equation}

Next, we consider an exemplary time-reversal procedure for a spreading single-particle wave packet with the quadratic spectrum. Let the packet at the time $t=0$ be localized at the origin and have the Lorentzian shape with the width $\xi_0$:
\begin{equation}
\Psi(x,0) = \sqrt{\frac{\xi_0}{2\pi}}\, \frac{2\xi_0}{x^2+\xi^2} \equiv \sum_p \sqrt{2\pi\xi_0} e^{-|p|\xi_0} e^{ipx}.
\end{equation}
A subsequent free evolution with quadratic Hamiltonian $\hat{H} = \hbar^2\hat{p}^2/2m$ during the time interval $\tau>0$ broadens the particle's wave function into
\begin{equation}
\Psi(x,\tau) \approx \frac{e^{-|x|m\xi_0/\hbar\tau}}{\sqrt{2\pi\hbar\tau/m}}\exp\Bigl( \frac{imx^2}{2\hbar\tau} \Bigr),
\end{equation}
having the typical size $\xi_\tau = \hbar\tau/m\xi_0$ or, equivalently, $\xi_\tau/\xi_0 = 4 \bar{E}\tau/\hbar$, where $\bar{E} = \hbar^2/4m\xi_0^2$ is the average energy carried by the wave-packet. The statistical sum $Z_\beta$ within the volume $\sim\xi_\tau$ is given by
\begin{equation}
      Z_\beta \sim \xi_\tau \int dE \nu_{\scriptscriptstyle\rm 1D} (E) e^{-\beta E} \sim \frac{\tau}{\hbar} \sqrt{\frac{\bar{E}}{\beta}},
\end{equation}
where $\nu_{\scriptscriptstyle 1D}(E) = (m/2\pi^2\hbar^2 E)^{1/2}$ is one-dimensional density of states. Assuming $E_\mathrm{max} \sim \bar{E}$, the reversal complexity for the time-reversal procedure with the accuracy $\epsilon$ is given by [see Eqs.(\ref{eq:temp}) and (\ref{eq:taccuracy})]
\begin{equation}
      N_\epsilon \sim \frac1{\epsilon^4} \Bigl( \frac{\bar{E}\tau}{\hbar} \Bigr)^7 = \frac1{\epsilon^4} \Bigl( \frac{\xi_\tau}{\xi_0} \Bigr)^7.
\end{equation}
The optimal inverse temperature of the thermostat is then given by
\begin{equation}
      \beta \sim \frac1{\bar{E}} \frac{\epsilon}{\xi_\tau/\xi_0}.
\end{equation}
Comparing this with the reversal complexity of a known state of the wave-packet, $N^\prime_\epsilon$$\sim$$\epsilon^{-1} (\xi_\tau/\xi_0)$, see Ref.\,[\onlinecite{us}], one finds that the reversal of an unknown wave-packet state is a more laborious computational task.

In conclusion, we have described the time-reversal procedure of an unknown mixed quantum state. The procedure relies on the ability to perform the LMR protocol and on the existence of an ancilla system whose dynamics is governed by the same Hamiltonian as the Hamiltonian of the reversed system,  which is not required to be known to us. The reversal procedure is comprised of $N\gg 1$ sequential applications of the LMR procedure
to the joint state of the system and ancilla prepared in a thermal state. In contrast to the known state reversal procedure, the introduced algorithm does not require to keep an information about all amplitudes of the reversed state. Yet the reversal complexity given by $N$ scales typically as squared dimension of a Hilbert space spanned the unknown state. Moreover, the operation rate of the LMR procedure has to be sufficiently high to overrun the forward time evolution of the reversed system during the execution of the reversal protocol. We expect that this protocol will be experimentally implemented on the few-qubit system in the nearest future and will become a subject of forthcoming publications.

\paragraph*{Acknowledgements.}
The work was supported by the U.S. Department of Energy, Office of Science, Basic Energy Sciences, Materials Sciences and Engineering Division (V.M.V. and partly A.V.L.) and by the RFBR Grant No. 18-02-00642A (A.V.L).  A.V.L. acknowledges the support from the Ministry of the Education and Science of the Russian Federation 16.7162.2017/8.9 and from the Government of the Russian Federation (Agreement 05.Y09.21.0018).

\appendix

\subsection{LMR procedure} \label{A1}

The LMR procedure goes as follows: one considers a combined system of the system in question and an ancilla $\hat\rho \otimes \hat\sigma$ and performs the joint unitary evolution over an infinitesimal time instant $\delta t$ under a Hamiltonian $\hbar\omega \hat{S}$,
\begin{equation}\label{eq:swapU}
\hat\rho \otimes \hat\sigma \to \exp\bigl(-i\omega\delta t \hat{S}\bigr) \bigl[ \hat\rho \otimes \hat\sigma \bigr] \exp \bigl( + i \omega\delta t \hat{S} \bigr),
\end{equation}
where $\hat{S}$ is a unitary SWAP operator\cite{SWAP} acting on the system and ancilla: $\hat{S}\bigl( |x\rangle_{\cal S} \otimes |y\rangle_a \bigr) = |y\rangle_{\cal S} \otimes |x\rangle_a$. The operator $\hat{S}$ is itself Hermitian and, therefore, the unitary operator $\exp(-i\omega\delta t \hat{S})$ can be implemented. Making use of its property $\hat{S}^2 = \hat{1}$, one gets $\exp(-i\omega\delta t \hat{S}) = \hat{1} \cos(\omega\delta t) -i \hat{S} \sin(\omega \delta t)$ and, therefore, its computational complexity is equivalent to the complexity of the unitary swap operator acting on the direct product of Hilbert spaces with the dimensions $\dim{\cal S}$. Next, we trace out the ancilla and get the quantum channel for the system's density matrix
\begin{eqnarray}\label{eq:Schannel}
&&\rho \to \Phi_{\delta t}[\hat\rho] =\cos^2(\omega\delta t) \hat\rho   + \sin^2(\omega \delta t) \hat\sigma
\\
&&\qquad -i\sin(\omega\delta t) \cos(\omega\delta t) \bigl[ \hat\sigma, \hat\rho \bigr].
\nonumber
\end{eqnarray}
At the infinitesimal time instant $\omega\delta t\to 0$ one gets the channel, $\Phi_{\delta t}\bigl[\hat\rho\bigr] = \hat\rho - i\omega\delta t \bigl[ \hat\sigma,\hat\rho \bigr]$, describing an infinitesimal time evolution under the Hamiltonian $\hbar\omega \hat\sigma$. Keeping $O\bigl( (\omega\delta t)^2\bigr)$ terms, one gets
\begin{eqnarray}
      &&\Phi_{\delta t}\bigl[ \hat\rho(t) \bigr] \approx e^{-i\omega\delta t \hat\sigma} \hat\rho(t) e^{+i\omega\delta t\hat\sigma}
      \\
      &&\qquad+(\omega\delta t)^2\Bigl( \hat\sigma + \frac12 \bigl[ \hat\sigma \bigl[ \hat\sigma, \hat\rho(t)\bigr]\bigr] -\hat\rho(t) \Bigr).
      \nonumber
\end{eqnarray}
Repeating the above procedure $N$ times one can generate the forward time evolution $\exp(-i\omega\tau \hat\sigma)$ over a finite time interval $\tau = N\delta t$
\begin{equation}\label{eq:rev_channel}
\Phi^N_{\delta t}[\hat\rho(t)] \approx \exp(-i\omega\tau \hat\sigma) \hat\rho \exp(+i\omega \tau \hat\sigma) + \delta\hat\rho(\tau),
\end{equation}
where the approximate accuracy is given by
\begin{equation}\label{eq:TRac1}
\delta\hat\rho(\tau) = \frac{(\omega\tau)^2}{N} \Bigl( \sigma + \frac12 \bigl[ \hat\sigma \bigl[ \hat\sigma, \hat\rho(t+\tau) \bigr] \bigr] - \hat\rho(t+\tau) \Bigr),
\end{equation}
with $\hat\rho(t+\tau) = \exp(-i\omega\tau \hat\sigma) \hat\rho(t) \exp(+i\omega\tau \hat\sigma)$ being the final state of the system.
\vspace{1cm}

\subsection{High entropy state reversal complexity} \label{A2}

Here we derive the Eq.\,(\ref{eq:hentN}) for the time-reversal complexity of the state $\hat\rho$ with the entropy $S = \ln\dim(N)-k\ln(2)$, where $N=\dim({\cal S})$ is the Hilbert space dimension of the system. The norm $||\hat\rho||$ is given by its maximum eigenvalue $||\hat\rho|| = p_1 > p_i$, $i=2,\dots N$ of the density operator. The von Neumann entropy can be decomposed into a sum
\begin{equation}\label{eq:S1}
      S = H(p_1) + (1-p_1) \sum_{i=2}^N \tilde{p}_i \ln(\tilde{p}_i),
\end{equation}
where $\tilde{p}_i = p_i/(1-p_1)$ with $\sum_{i=2}^N \tilde{p}_i = 1$, $H(x) = -x\ln(x) -(1-x)\ln(1-x)\leq \ln(2)$. Let us find a maximal possible $p_1$ for a given $S$. One sees straightforwardly that $p_1$ is maximal if all $\tilde{p}_{i}$, $i=2,\dots N$ are uniform and Eq.(\ref{eq:S1}) is reduced to
\begin{equation}
      \ln(N) -k\ln(2) = H(p_1) +(1-p_1) \ln(N-1).
\end{equation}
For $N$$\gg$$1$ one can assume $p_1$$\ll$$1$ and get the approximate solution $p_1$$\approx$$k/\log_2(N)$ that results in Eq.\,(\ref{eq:hentN}).

\end{document}